\documentclass[twocolumn, 10pt, pre, aps, showpacs, reprint, amsmath, amssymb, superscriptaddress]{revtex4-1}

\usepackage[pdftex]{graphicx}
\usepackage{subfigure}

\begin{document}

\newcommand{\kt}{\gamma}
\newcommand{\lzt}{q_z}
\newcommand{\atled}{\bm{\nabla}}
\newcommand{\dx}{\frac{\partial}{\partial_x}}
\newcommand{\dy}{\frac{\partial}{\partial_y}}
\newcommand{\dz}{\frac{\partial}{\partial_z}}
\newcommand{\dt}{\frac{\partial}{\partial_t}}
\newcommand{\sqrdt}{\frac{\partial^2}{\partial_t^2}}
\newcommand{\pbyp}[2]{\frac{\partial #1}{\partial #2}}
\newcommand{\dbyd}[2]{\frac{d #1}{d #2}}
\newcommand{\ex}{\bm{e}_x}
\newcommand{\ey}{\bm{e}_y}
\newcommand{\ez}{\bm{e}_z}
\newcommand{\besselj}[2]{\mathrm{J}_{#1}(#2)}
\newcommand{\besseljp}[2]{\mathrm{J'}_{#1}(#2)}
\newcommand{\hankel}[3]{\mathrm{H}_{#1}^{(#2)}(#3)}
\newcommand{\hankelp}[3]{\mathrm{H'}_{#1}^{(#2)}(#3)}
\newcommand{\laplace}{\Delta}
\newcommand{\neff}{\tilde{n}}
\newcommand{\fexp}{f_{\mathrm{expt}}}
\newcommand{\ftheo}{f_{\mathrm{calc}}}
\newcommand{\nexp}{\tilde{n}}
\newcommand{\Gtheo}{\Gamma_{\mathrm{calc}}}
\newcommand{\Gexp}{\Gamma_{\mathrm{expt}}}
\newcommand{\Grad}{\Gamma_{\mathrm{rad}}}
\newcommand{\Gabs}{\Gamma_{\mathrm{abs}}}
\newcommand{\Gant}{\Gamma_{\mathrm{ant}}}
\newcommand{\rhof}{\rho_{\mathrm{fluc}}}
\newcommand{\rhofscl}{\rhof^{\mathrm{scl}}}
\newcommand{\rhofSS}{\rhof^{(\mathrm{ss})}}
\newcommand{\rhofnr}[1]{\rho_{#1 n_r}}
\newcommand{\rhow}{\rho_{\mathrm{Weyl}}}
\newcommand{\Nweyl}{N_{\mathrm{Weyl}}}
\newcommand{\rhot}{\tilde{\rho}}
\newcommand{\rhotscl}{\rhot_{\mathrm{scl}}}
\newcommand{\rhotSS}{\rhot^{(\mathrm{ss})}}
\newcommand{\rhotnr}[1]{\tilde{\rho}_{#1 n_r}}
\newcommand{\kmin}{k_{\mathrm{min}}}
\newcommand{\kmax}{k_{\mathrm{max}}}
\newcommand{\fmax}{f_{\mathrm{max}}}
\newcommand{\po}{\mathrm{po}}
\newcommand{\lpo}{\ell_\po}
\newcommand{\lmax}{\ell_{\mathrm{max}}}
\newcommand{\alphacrit}{\alpha_{\mathrm{crit}}}
\newcommand{\chico}{\chi_{\mathrm{co}}}
\newcommand{\reffig}[1]{\mbox{Fig. \ref{#1}}}
\newcommand{\subreffig}[1]{\mbox{Fig. \subref{#1}}}
\newcommand{\refeq}[1]{\mbox{Eq. (\ref{#1})}}
\newcommand{\refsec}[1]{\mbox{Sec.\ \ref{#1}}}
\newcommand{\reftab}[1]{\mbox{Table \ref{#1}}}
\newcommand{\etal}{\textit{et al.\ }}
\renewcommand{\Re}[1]{\mathrm{Re}(#1)}
\renewcommand{\Im}[1]{\mathrm{Im}(#1)}

\hyphenation{re-so-nan-ce re-so-nan-ces ex-ci-ta-tion z-ex-ci-ta-tion di-elec-tric ap-pro-xi-ma-tion ra-dia-tion Me-cha-nics quan-tum pro-posed Con-cepts pro-duct Reh-feld ob-ser-va-ble Se-ve-ral rea-so-nable Ap-pa-rent-ly re-pe-ti-tions re-la-tive quan-tum su-per-con-duc-ting ap-pro-xi-mate cri-ti-cal}

\title{Experimental test of a trace formula for two-dimensional dielectric resonators}

\author{S. Bittner}
\affiliation{Institut f\"ur Kernphysik, Technische Universit\"at Darmstadt, D-64289 Darmstadt, Germany}
\author{E. Bogomolny}
\affiliation{Laboratoire de Physique Th{\'e}orique et Mod{\`e}les Statistiques, CNRS-Universit{\'e} Paris-Sud, 91405 Orsay, France}
\author{B. Dietz}
\author{M. Miski-Oglu}
\author{P. Oria Iriarte}
\affiliation{Institut f\"ur Kernphysik, Technische Universit\"at Darmstadt, D-64289 Darmstadt, Germany}
\author{A. Richter}
\email{richter@ikp.tu-darmstadt.de}
\affiliation{Institut f\"ur Kernphysik, Technische Universit\"at Darmstadt, D-64289 Darmstadt, Germany}
\affiliation{ECT*, Villa Tambosi, Villazano, I-38123 Trento, Italy}
\author{F. Sch\"afer}
\affiliation{LENS, University of Florence, Sesto-Fiorentino, I-50019 Firenze, Italy}

\date{\today}

\begin{abstract}
Resonance spectra of two-dimensional dielectric microwave resonators of circular and square shapes have been measured. The deduced length spectra of periodic orbits were analyzed and a trace formula for dielectric resonators recently proposed by Bogomolny \etal [Phys.\ Rev.\ E \textbf{78}, 056202 (2008)] was tested. The observed deviations between the experimental length spectra and the predictions of the trace formula are attributed to a large number of missing resonances in the measured spectra. We show that by taking into account the systematics of observed and missing resonances the experimental length spectra are fully understood. In particular, a connection between the most long-lived resonances and certain periodic orbits is established experimentally.

\end{abstract}


\maketitle

\section{\label{sec:intr}Introduction}
Trace formulas relate the density of states of a quantum system to the periodic orbits (POs) of the corresponding classical system. They were first introduced by Gutzwiller \cite{Gutzwiller1970, Gutzwiller1971, Gutzwiller1990} and have since then found numerous applications not only to quantum systems but also to wave-dynamical systems such as electromagnetic \cite{Balian1977, Dembowski2002} or acoustic \cite{Wirzba2005} resonators. In addition concepts from quantum chaos and semiclassics are nowadays also applied to open dielectric resonators (also called dielectric billiards), which are used, e.g., as microlasers, as sensors or in optical circuits \cite{Noeckel2002, Vahala2004}. Especially the occurrence of scarred resonance states \cite{Gmachl2002, Harayama2003} and the connections between the emission properties of microlasers and POs have gained considerable attention. It has been established that the directions of maximal emission from a microlaser with, e.g., quadrupole shape are determined by the unstable manifolds of certain POs \cite{Schwefel2004, Schaefer2006}. Recently, a trace formula for two-dimensional (2D) dielectric billiards was proposed in \cite{Bogomolny2008}. This trace formula provides a connection between the resonance density of a dielectric resonator as a wave-dynamical system and the POs of the corresponding classical ray-dynamical billiard system and is a continuation of ideas developed in \cite{Fukushima2006} and \cite{Lebental2007}. \newline
The objective of the present work is an experimental test of the trace formula \cite{Bogomolny2008} with 2D dielectric resonators. Due to radiation losses, the resonances of a dielectric billiard have finite lifetimes. The lifetimes $\tau_j$ depend on the individual resonances, and the corresponding resonance widths $\Gamma_j = 1 / \tau_j$ range over several orders of magnitude. Therefore, the spectrum consists of a mixture of resonances with very small up to extremely large widths, and only the sharp, long-lived resonances can be clearly identified. Thus, only a part of the resonances is actually observed in an experiment. This is an important difference to hardwalled billiards, for which complete spectra can be measured using superconducting microwave resonators \cite{Richter1999, Dembowski2001, Dembowski2002}. The task of the present work is to demonstrate that nevertheless the measured, incomplete spectrum can still be interpreted by means of the trace formula. Furthermore, long-lived resonances, which play an important role in microlasers, and their relation to certain POs are investigated. It should be noted that flat microlasers are usually approximated as 2D systems by introducing the so-called effective index of refraction, even though the precision of this approximation is not always under control \cite{Bittner2009}. The applicability of the trace formula on such flat three-dimensional systems within this 2D approximation will be the subject of a future publication. \newline
Microwave resonators provide a suitable testbed for the investigation of dielectric resonators because of their macroscopic dimensions and the large spectral range accessible experimentally. Here we use flat dielectric plates of different shapes and materials as passive resonators and put them between two copper plates. Then, up to a certain frequency, they are described by a 2D scalar Helmholtz equation \cite{Richter1999}. Three different resonators with regular classical dynamics were investigated, a circular and a square one made of Teflon each with index of refraction $n \approx 1.4$ and a square resonator made of alumina (Al$_2$O$_3$) with $n \approx 3$. The paper is organized as follows. Details of the trace formula are briefly summarized in \refsec{sec:trform} and the experimental setup is described in \refsec{sec:expset}. The results for the circular Teflon resonator, the square Teflon resonator, and the square alumina resonator are presented in Secs.\ \ref{sec:tefloncircle}, \ref{sec:teflonsquare}, and \ref{sec:aluminasquare}, respectively. Section \ref{sec:conc} concludes with a discussion of the results and a summary. \newline

\section{\label{sec:trform}Trace formula for dielectric resonators}
As already noted above, trace formulas relate the density of states of a wave-dynamical system to a sum over all POs of the corresponding classical ray-dynamical system. An open 2D dielectric resonator is a flat cylinder whose cross-sectional area has an arbitrary shape made of a dielectric material with index of refraction $n > 1$ surrounded by air (or another material with lower index of refraction). The corresponding classical system is the billiard with the same shape. Rays travel freely inside the billiard domain $S$ and are partially reflected back inside and partially transmitted outside of the billiard according to the Fresnel formulas \cite{Hentschel2009} when they hit the boundary $\partial S$. The wave equations used for such 2D dielectric resonators are \cite{Jackson1999}
\begin{equation} \begin{array}{c} ( \laplace + n^2 k^2 ) \, \Psi(\vec{r}) \, , \, \vec{r} \, \mathrm{inside} \, S, \\ 
( \laplace + k^2 ) \, \Psi(\vec{r}) \, , \, \vec{r} \, \mathrm{outside} \, S. \end{array} \end{equation}
The wave number $k$ is related to the frequency $f$ via $k = 2 \pi f / c$, where $c$ is the speed of light in vacuum. Below a certain frequency only transverse magnetic (TM) field modes are excited in the resonator setup considered here (see \refsec{sec:expset}). Accordingly, the wave function $\Psi$ corresponds to the $z$-component of the electric field, $E_z$, and both the wave function and its normal derivative are continuous along the boundary $\partial S$. In such a resonator there are only quasi-bound states or resonances, which are characterized by complex frequencies $f_j$, where $\Re{f_j}$ is the resonance frequency and $\Gamma_j = -2 \, \Im{f_j}$ is the resonance width (full width at half maximum) \cite{Noeckel2002}. The resonance frequencies and widths are obtained by fitting Lorentzians to the measured frequency spectra. The spectral density of states is defined as 
\begin{equation} \label{eq:dos} \rho(k) = - \frac{1}{\pi} \sum \limits_j \frac{\Im{k_j}}{[k - \Re{k_j}]^2 + [\Im{k_j}]^2} \, . \end{equation}
The density of states can generally be decomposed into a smooth part (also known as Weyl term \cite{Brack2003}) and a fluctuating part, $\rho = \rhow + \rhof$, where the smooth part is related to the area $A$ and perimeter $U$ of the resonator and the fluctuating part to the POs of the classical billiard. For a 2D dielectric resonator the smooth part of the density of states is \cite{Bogomolny2008}
\begin{equation} \label{eq:rho_weyl} \rhow(k) = \frac{A n^2}{2 \pi}k + \tilde{r}(n) \frac{U}{4 \pi} \, . \end{equation}
Here, $\tilde{r}$ is related to the boundary conditions. It is defined as
\begin{equation} \tilde{r}(n) = 1 + \frac{n^2}{\pi} \int \limits_{-\infty}^{\infty} \frac{dt}{t^2 + n^2}\tilde{R}(t) - \frac{1}{\pi} \int \limits_{-\infty}^{\infty} \frac{dt}{t^2 + 1}\tilde{R}(t) \, , \end{equation}
where for TM modes
\begin{equation} \tilde{R}(t) = \frac{\sqrt{t^2 + n^2} - \sqrt{t^2 + 1}}{\sqrt{t^2 + n^2} + \sqrt{t^2 + 1}} \, . \end{equation}
In the case of a regular classical billiard the semiclassical expression for $\rhof$ reads as \cite{Bogomolny2008}
\begin{equation} \label{eq:rho_scl} \rhofscl(k) = \sum \limits_{\po} \sqrt{\frac{n^3}{\pi^3}} B_\po |R_\po| \sqrt{k} \, e^{i(n k \lpo + \varphi_\po)} + \mathrm{c.c.} \end{equation}
Here, $B_\po \propto A_\po / \sqrt{\lpo}$, where $A_\po$ is the area of the billiard in configuration space covered by the family of POs with length $\lpo$, $R_\po$ denotes the product of all Fresnel reflection coefficients resulting from reflections at the boundary of the billiard, and $\varphi_\po$ denotes the phase changes accumulated at the reflections [i.e., $\arg{(R_\po)}$] and the caustic points \cite{Bogomolny2008, Brack2003}. The details of $\lpo$, $B_\po$, $\varphi_\po$ and $R_\po$ for the dielectric circle and square billiards are given in Appendixes \ref{sec:po_circ} and \ref{sec:po_square}. For $\tilde{r} = -1$ this trace formula equals that of a closed 2D quantum billiard with Dirichlet boundary conditions except for some additional terms involving the index of refraction $n$ which reflect the larger optical length or volume of the resonator and the contributions of the Fresnel coefficients. The semiclassical expression $\rhow + \rhofscl$ is a good approximation to the density of states $\rho(k)$ defined by \refeq{eq:dos} in the semiclassical limit $k \rightarrow \infty$. The Fourier transform (FT) of $\rhofscl(k)$, 
\begin{equation} \label{eq:lspec_scl} \rhotscl(\ell) = \int_{\kmin}^{\kmax} dk \, \rhofscl(k) \, e^{-i k n \ell} \, , \end{equation}
yields the length spectrum which has peaks at the lengths $\lpo$ of the periodic orbits. For a test of the trace formula we compare it with the FT of the fluctuating part of the density of states,
\begin{equation} \label{eq:lspec} \begin{array}{rcl} \rhot(\ell) & = & \int_{\kmin}^{\kmax} dk \left[ \rho(k) - \rhow(k) \right] e^{-i k n \ell} \\ \\
 & = & \sum \limits_{j} e^{-i k_j n \ell} - \mathrm{FT}\{ \rhow \} \, . \end{array} \end{equation}
The summation in \refeq{eq:lspec} is over \textit{all} resonances with $\kmin \leq \Re{k_j} \leq \kmax$. Short-lived resonances are suppressed because of the factor $\exp{(-n l |\Im{k_j}|)}$ appearing in \refeq{eq:lspec} as compared to long-lived resonances.

\section{\label{sec:expset}Experimental setup}

\begin{figure}[tb]
\subfigure[]{
	\includegraphics[width = 8.4 cm]{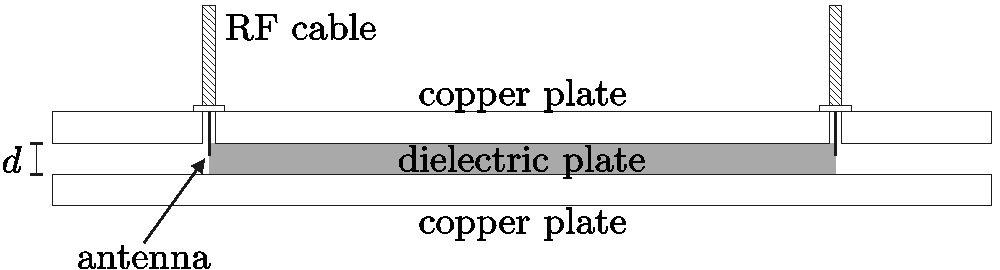}
	\label{subfig:setupside}	
}
\subfigure[]{
	\includegraphics[width = 8.4 cm]{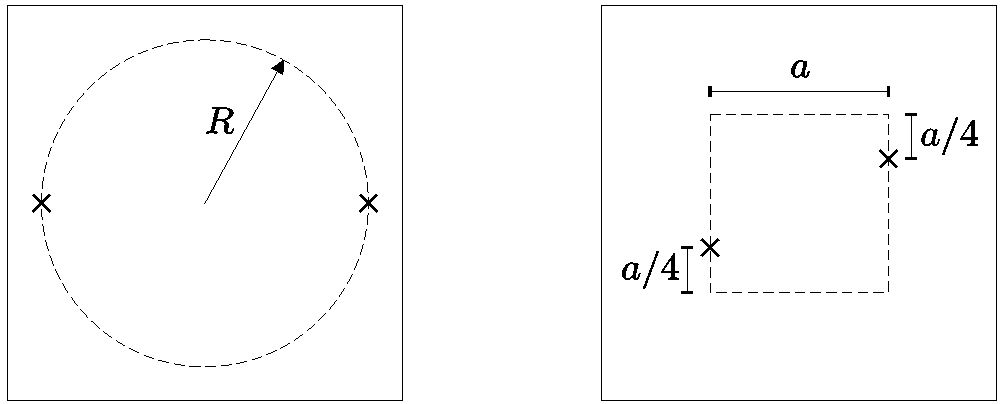}
	\label{subfig:setuptop}	
}
\caption{\label{fig:setup}Schematic picture of the experimental setup (not to scale). \subref{subfig:setupside} Side view: the dielectric plate is placed between two copper plates. Two dipole antennas entering the resonator through small holes in the top plate are placed next to the sidewalls of the dielectric plate. The dipole antennas are attached to a vectorial network analyzer via rf cables. \subref{subfig:setuptop} Top view: the solid lines denote the contour of the copper plates; the dashed lines denote that of the dielectric plates (circle with radius $R$ and squares with side length $a$). The crosses indicate the positions of the antennas.}
\end{figure}

\begin{figure*}[t]
\includegraphics[width = 16 cm]{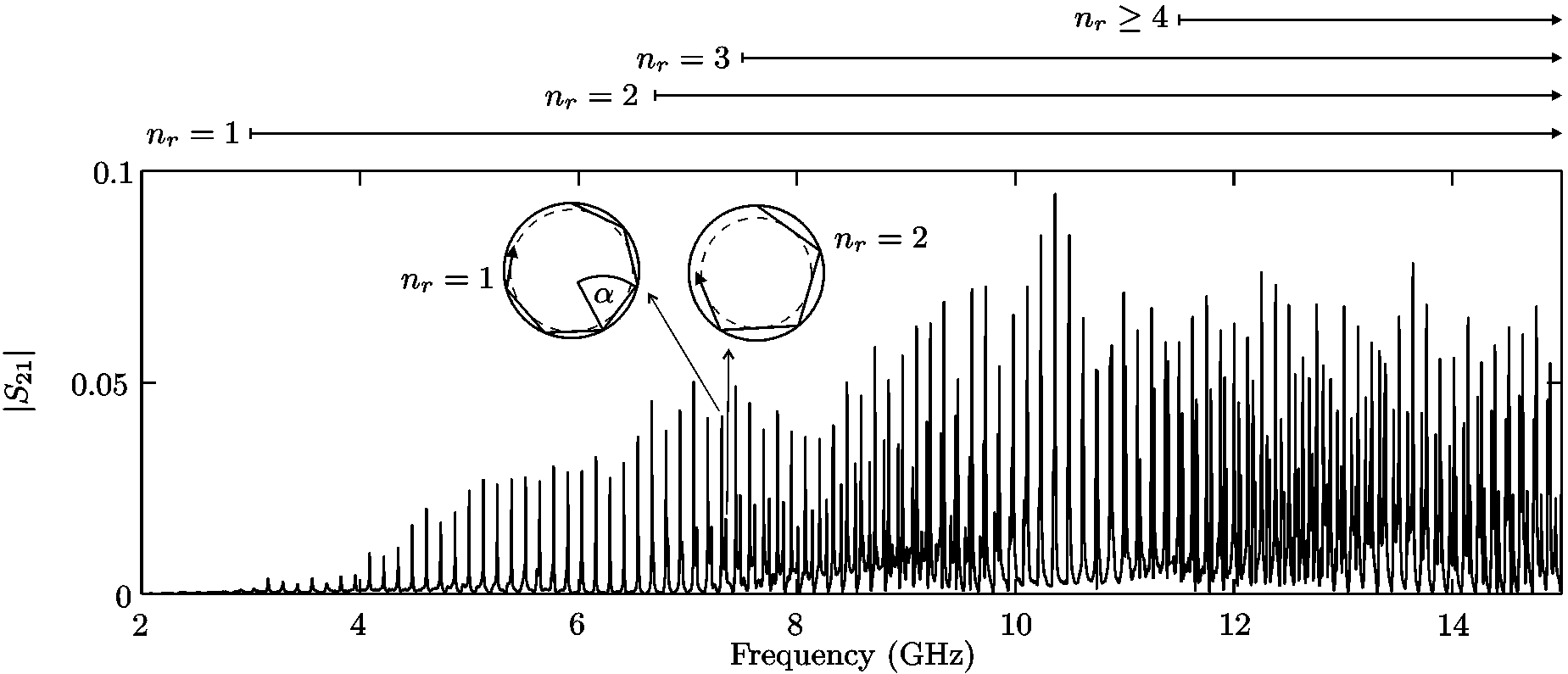}
\caption{\label{fig:spec_Tefcirc}Frequency spectrum of the Teflon circle. The modulus of the transmission amplitude $S_{21}$ is plotted with respect to the frequency $f$. The classical trajectories associated with two resonances with $n_r = 1$ and $n_r = 2$ are shown as insets, and $\alpha$ denotes the angle of incidence with respect to the surface normal of these trajectories. The bars above the graph denote the frequency regimes in which modes with certain radial quantum numbers $n_r$ are observed.}
\end{figure*}

A sketch of the experimental setup is shown in \reffig{fig:setup}: a plate made of a dielectric material (Teflon or alumina) is placed between two copper plates. Then below the frequency
\begin{equation} f_{\mathrm{2D}} = \frac{c}{2 n d} \, , \end{equation}
only TM modes with homogeneous electric field in the $z$-direction ($\mathrm{TM}_0$) exist. Here, $d$ is the thickness and $n$ is the index of refraction of the dielectric plate. Below this frequency, the resonator can be treated exactly as a 2D system \cite{Richter1999}. Microwave power is coupled into and out of the resonator with two dipole antennas. A vectorial network analyzer (PNA 5230A by Agilent Technologies) is used to measure the complex transmission amplitude $S_{21}(f)$, where the modulus squared of $S_{21}$ equals the ratio
\begin{equation} |S_{21}(f)|^2 = \frac{P_{\mathrm{out}}}{P_{\mathrm{in}}} \end{equation}
between the power $P_{\mathrm{in}}$ coupled in by antenna $1$ and the power $P_{\mathrm{out}}$ coupled out via antenna $2$ for a given frequency $f$.
The dipole antennas are put next to the sidewalls of the dielectric plates [see \reffig{subfig:setupside}], so that they can couple to the evanescent fields of the resonance modes. Their positions are indicated in \reffig{subfig:setuptop}. The measured transmission amplitude $|S_{21}|$ at a resonance frequency rises with the electric field strengths $E_z$ at the positions of the two antennas \cite{Stein1995}. In the case of the circle billiard, $E_z \propto \cos{(m  \varphi)}$ with $\varphi$ and $m$ being the azimuthal angle and quantum number, respectively, so two antennas placed on opposite sides of the billiard ensure optimal coupling to all resonances. In the case of the square billiards, the antennas were placed offside any symmetry axes so that they couple to resonance states of all symmetry classes. It was checked that the results presented here do not depend on the specific positions. The influence of waves reflected at the edges of the copper plates can be neglected. Details concerning the three different dielectric billiards are given in the corresponding sections. Since the indices of refraction of the dielectric plates are only known with an uncertainty of a few percent, the precise values of $n$ of the different plates were deduced from the length spectra by adjusting $n$ such that the positions of the peaks in the length spectra match the lengths of the corresponding POs \cite{Note1}, as will be shown below. The quality factors $Q_j = \Re{f_j} / \Gamma_j$ of the measured resonances are always smaller than those expected theoretically because there are Ohmic losses in the copper plates and the antennas and absorption in the dielectric material in addition to the pure radiation losses.

\section{\label{sec:tefloncircle}Circular Teflon resonator}

The first resonator investigated is a circular disk made of Teflon (Gr{\"u}nberg Kunststoffe GmbH). The resonator (called the Teflon circle in the following) has a radius of \mbox{$R = 274.9$ mm} and a thickness of \mbox{$d = 5.0$ mm}. Its index of refraction deduced from the length spectrum is $n = 1.419 \pm 0.001$. Therefore, the critical angle for total internal reflection (TIR) is $\alphacrit = 44.8^\circ$. The dipole antennas were placed along the diameter of the disk on opposite sides [see \reffig{subfig:setuptop}]. A measured frequency spectrum is shown in \reffig{fig:spec_Tefcirc}, with a frequency of \mbox{$10$ GHz} corresponding to $kR = 57.6$. The spectrum features several families of almost equidistant, sharp resonances. These can be labeled with azimuthal and radial quantum numbers $(m, n_r)$, and each family consists of resonances with the same $n_r$ and different $m$. The quality factors of the resonances are typically $Q = 1000 \-- 5000$. Since the radiation losses $|\Im{f_{m, n_r}}|$ increase with increasing $n_r$, only modes with small $n_r$ are observed in the measured spectrum. Modes with higher $n_r$ are visible at higher frequencies (compare, e.g., \cite{Bittner2009}). An angular momentum of $\hbar m$ can be attributed to each resonance. In the ray-picture, a trajectory with this angular momentum has an angle of incidence $\alpha$ with respect to the surface normal given by \cite{Hentschel2002b}
\begin{equation} \sin{\alpha} = \frac{m}{n \Re{k_j} R} \, . \end{equation}
These trajectories are in general not POs. Two examples of trajectories associated with a resonance with $n_r = 1$ and with one with $n_r = 2$ are shown as insets in \reffig{fig:spec_Tefcirc}. These show that a larger $n_r$ corresponds to a smaller angle of incidence $\alpha$, because for a given resonance frequency [i.e., $\Re{k_j}$], a mode with higher radial quantum number $n_r$ has a smaller azimuthal quantum number $m$, and a smaller angle of incidence results in larger radiation losses. The trajectories that are located close to the boundary of the circle have a large caustic (dashed inner circle) and are therefore called whispering gallery modes (WGMs). Indeed, all the long-lived resonances in the measured spectrum are of the WGM type. The bars above \reffig{fig:spec_Tefcirc} indicate the frequency regimes in which modes with different radial quantum numbers can be observed. Most of the observed resonances have $n_r = 1 \-- 3$ and only a few have $n_r \geq 4$. 

\begin{figure}[tb]
\includegraphics[width = 8.6 cm]{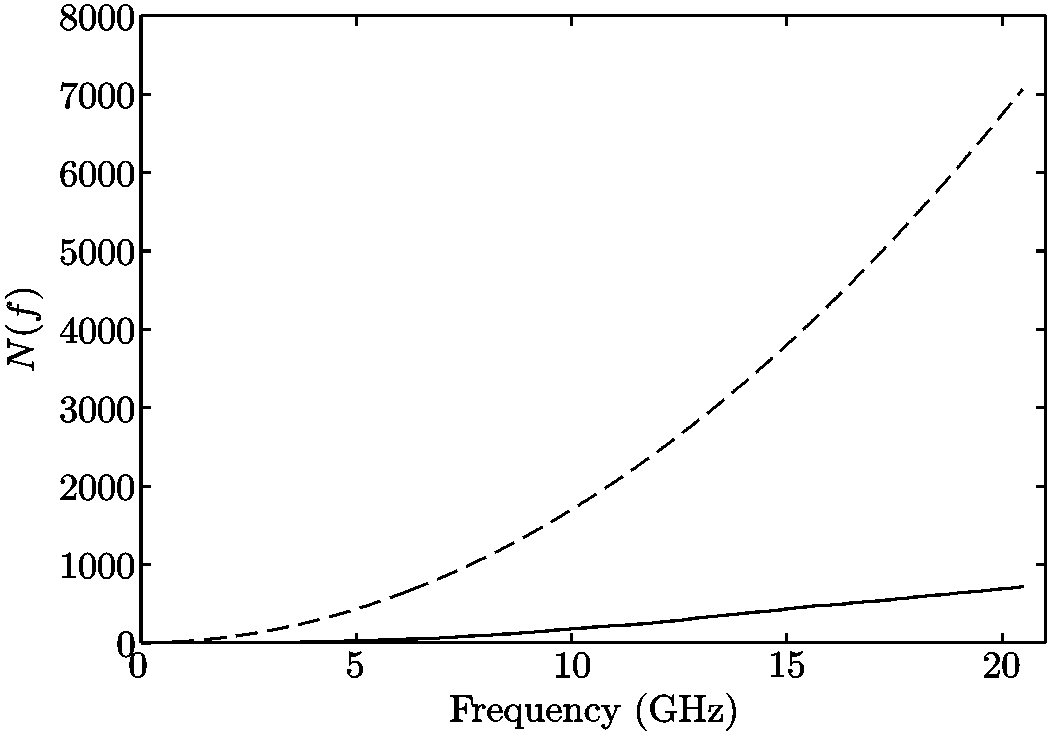}
\caption{\label{fig:staircase}The integrated resonance density $N(f)$ for the Teflon circle. The solid line is deduced from the measured frequency spectrum in \reffig{fig:spec_Tefcirc}; the dashed line is deduced from Weyl's law. All in all, $716$ resonances were identified in the spectrum up to \mbox{$20.5$ GHz}, which is only about $10 \, \%$ of the total number of resonances.}
\end{figure}

\begin{figure}[bt]
\includegraphics[width = 8.6 cm]{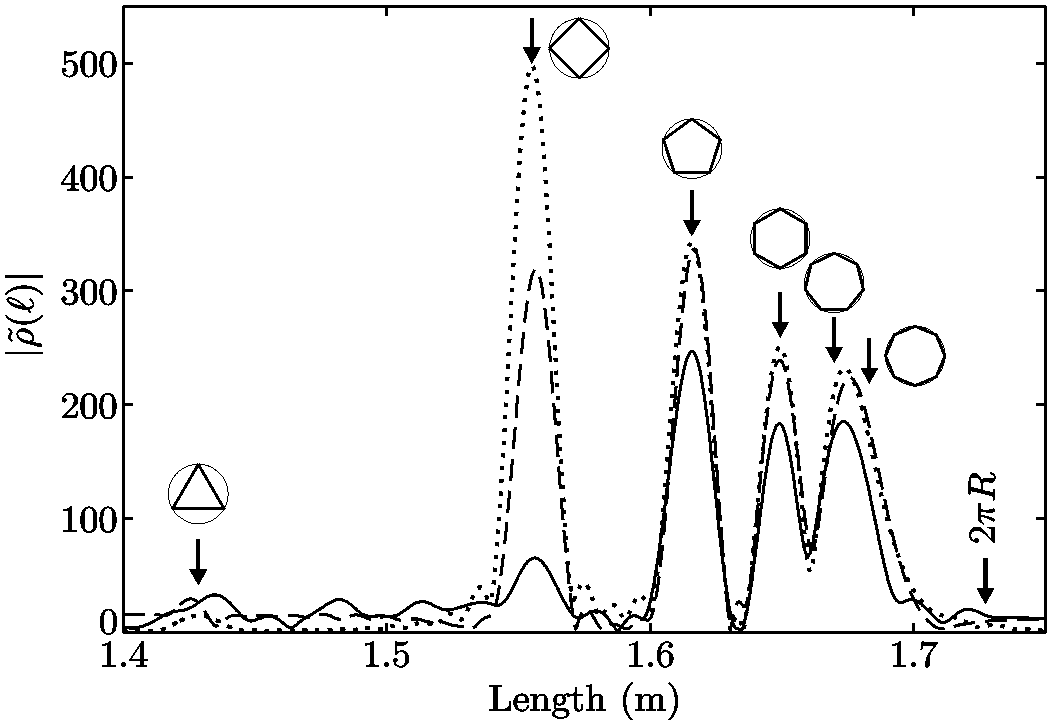}
\caption{\label{fig:Lspekt_Tefcirc_all}Length spectrum for the Teflon circle. The full line results from the measured spectrum, the dashed line results from a complete, calculated spectrum, and the dotted line shows the semiclassical expression $|\rhotscl(\ell)|$. The arrows indicate the lengths of the depicted POs and of the circumference $2 \pi R$ of the circle. The semiclassical expression and the calculated length spectrum agree well except for the square orbit. The experimental length spectrum has smaller amplitudes than the calculated one. However, the deviations are smaller for the higher-order polygon orbits.}
\end{figure}

In \reffig{fig:staircase} the integrated resonance density $N(f)$, which denotes the number of resonances below a given frequency $f$, is shown. The resonances observed in the measured spectrum (solid line) were counted twice because all modes with $m > 0$ are twofold degenerate. The dashed line is obtained from Weyl's law [\refeq{eq:rho_weyl}]. Only modes up to $20.5 \, \mathrm{GHz} \lesssim f_{\mathrm{2D}}$ are considered in the following. The comparison of Weyl's law and the integrated resonance density in \reffig{fig:staircase} illustrates that the long-lived resonances yield only a small part of the whole spectrum. \newline
The length spectrum for the Teflon circle is shown in \reffig{fig:Lspekt_Tefcirc_all}. The full line is the length spectrum obtained from the measured frequency spectrum via \refeq{eq:lspec}, the dashed line corresponds to the complete spectrum calculated by solving the Helmholtz equation for the dielectric circle \cite{Hentschel2002b}, and the dotted line is the semiclassical expression $|\rhotscl(\ell)|$ from \refeq{eq:lspec_scl}. All Fourier transforms were performed using the Welch-function as window function to smooth the resulting curves \cite{Press1987}. The positions of the peaks of the experimental length spectrum depend sensitively on the index of refraction $n$ used in the FT [cf.\ the term $\exp{(-i k n \ell)}$ in \refeq{eq:lspec}]. The positions of the peaks only coincide with the lengths of the POs if the correct value of $n$ is used, and thus $n$ is determined by matching these. The lengths of the different POs and the circumference are indicated by the arrows. The POs in the circle billiard have polygon and star shapes and are characterized by their periods and rotation numbers $(q, \eta)$, where $q$ is the number of reflections at the boundary and $\eta$ is the number of turns around the center, e.g., the $(4, 1)$ orbit is a square and the $(5, 2)$ orbit a pentagram. For the lengths considered in \reffig{fig:Lspekt_Tefcirc_all}, only polygonal POs exist ($\eta = 1$). The orbits with $q > 8$ are not indicated in the figure because their amplitudes $B_\po$ decrease rapidly with increasing $q$ (see Appendix \ref{sec:po_circ}). The semiclassical expression (dotted line) and the calculated length spectrum (dashed line) agree very well except for the case of the square orbit. Its angle of incidence $\alpha = 45^\circ$ is close to the critical angle for TIR. Thus, further corrections for $R_\po$ must be taken into account \cite{Bogomolny2008}. No discernible peaks are visible for orbits with angles of incidence smaller than the critical angle. The overall shape of the experimental length spectrum is reproduced by the calculated length spectrum, but it has smaller peak amplitudes, as to be expected due to the large number of missing resonances. In fact, in some cases the amplitudes of the peaks in the experimental length spectrum are as large as $80 \%$ of those in the calculated one. Thus, indeed the $10 \%$ most long-lived modes suffice to reproduce most of the peaks expected semiclassically. Interestingly, the agreement between the experimental and the calculated length spectra is better for the higher-order polygon orbits such as the hexagon and the heptagon, and worst for the square orbit. This might be explained by the fact that the experimental spectrum consists only of WGMs (see \reffig{fig:spec_Tefcirc}).

\begin{figure}[tb]
\includegraphics[width = 8.6 cm]{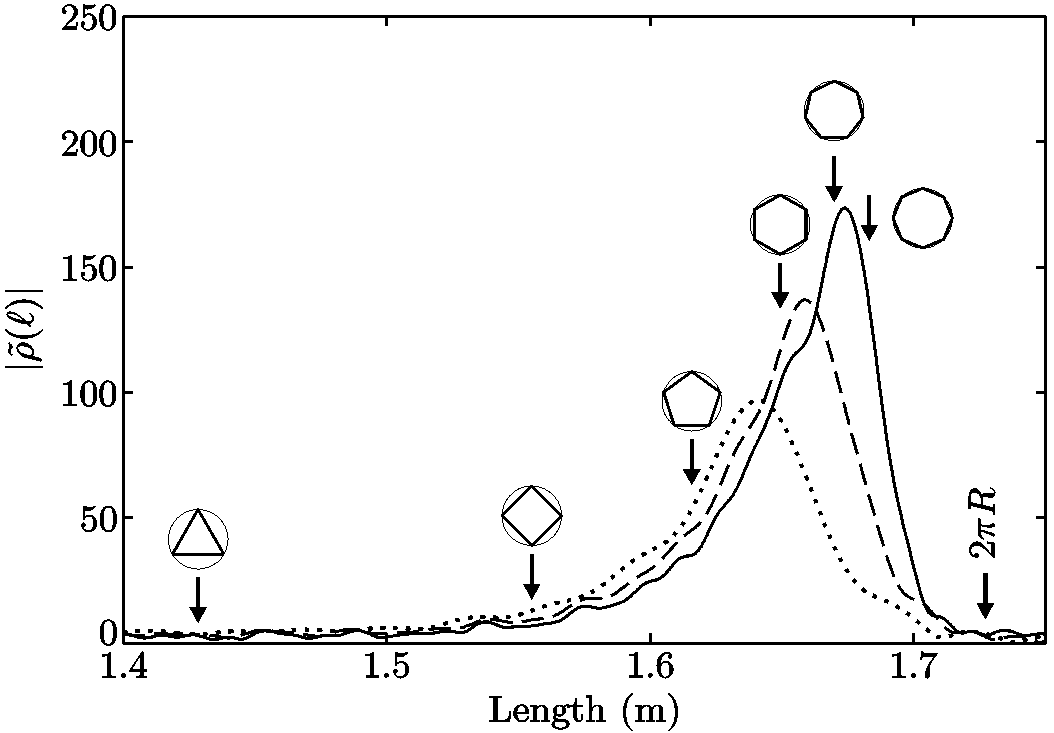}
\caption{\label{fig:Lspekt_Tefcirc_nr1}Length spectrum for a single resonance family. Only modes with radial quantum number $n_r = 1$ were considered and only up to a certain frequency $\fmax$ (solid line: \mbox{$20.5$ GHz}; dashed line: \mbox{$15$ GHz}; dotted line: \mbox{$10$ GHz}). The arrows denote the lengths of some POs and of the circumference $2 \pi R$ of the circle. Apparently, the position of the maximum of $|\rhot(\ell)|$ only depends on the frequency $\fmax$, and is not related to any PO.}
\end{figure}

\begin{figure}[tb]
\subfigure[]{
	\includegraphics[width = 8.4 cm]{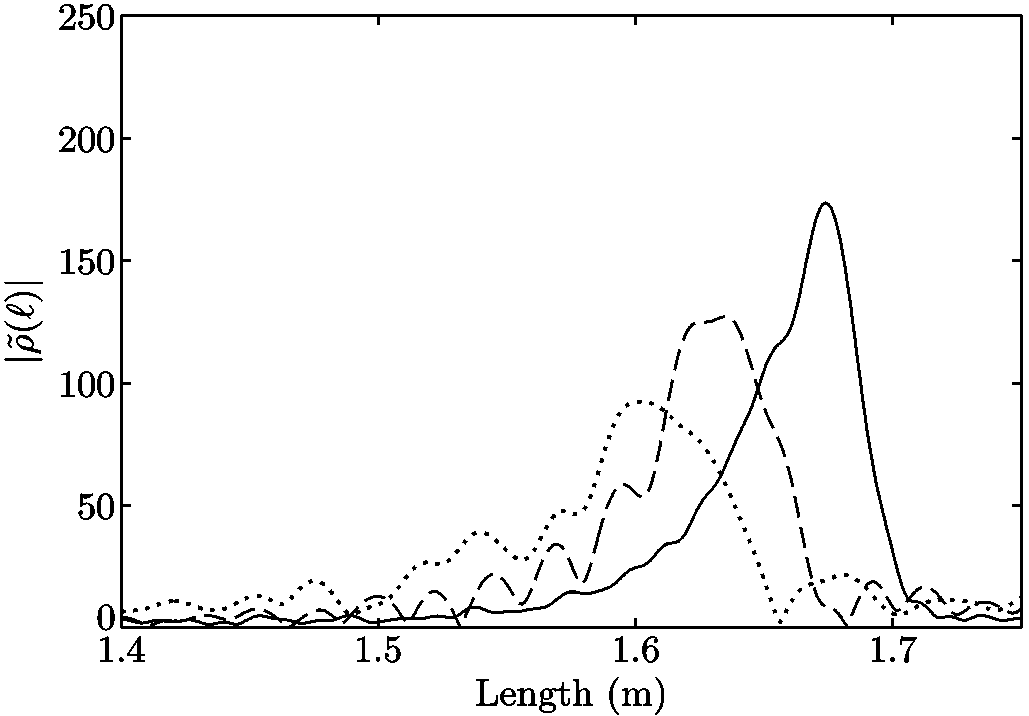}
	\label{subfig:Lspekt_Tefcirc_single_nr}
}
\subfigure[]{
	\includegraphics[width = 8.4 cm]{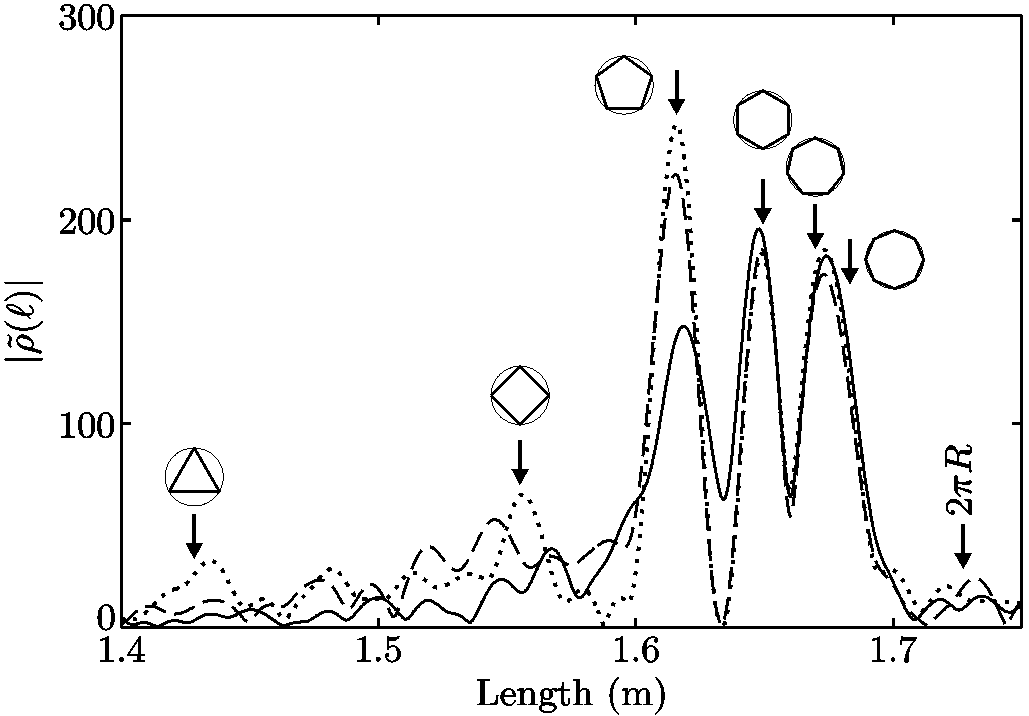}
	\label{subfig:Lspekt_Tefcirc_several_nr}
}
\caption{\label{fig:Lspekt_Tefcirc_nrs}Length spectra for different resonance families. \subref{subfig:Lspekt_Tefcirc_single_nr} Only one radial quantum number taken into account each (solid line: $n_r = 1$; dashed line: $n_r = 2$; dotted line: $n_r = 3$). \subref{subfig:Lspekt_Tefcirc_several_nr} Several families combined. The solid line is the length spectrum for the families with radial quantum numbers $n_r = 1$ and $2$ combined, the dashed line is for $n_r = 1, 2,$ and $3$ combined, and the dotted line is for all resonances (identical with the solid line in \reffig{fig:Lspekt_Tefcirc_all}). The peaks of the length spectra in \subref{subfig:Lspekt_Tefcirc_several_nr} originate from the interference between the different resonance families, and the main contributions of the different families are focused on different orbits.}
\end{figure}

To achieve a better understanding of the correspondence between different families of resonances and the different POs in the length spectrum, the measured spectrum was divided into subspectra with radial quantum numbers $n_r = 1, 2, 3$ and $n_r \geq 4$. The radial quantum numbers were identified by comparison with the calculated spectrum and by following the different series in the spectrum. First, only modes with $n_r = 1$ are considered, which form a family of almost equidistant resonances. Note that spectra containing only one such family are often encountered in microlaser and microcavity applications (e.g., \cite{Lebental2007, Fang2007a}). The length spectrum for the modes with $n_r = 1$ is depicted in \reffig{fig:Lspekt_Tefcirc_nr1}. Only resonances up to a certain frequency $\fmax$ were considered for the three different curves: the solid line shows the length spectrum for all modes with $n_r = 1$, the dashed line shows that for modes up to \mbox{$15$ GHz}, and the dotted line shows that for modes up to \mbox{$10$ GHz}. Each curve has only one peak, and, as expected, the position of the peak is close to the lengths of the high-order polygons, but the position of the peak does not correspond to any particular PO, instead it only depends on $\fmax$. This can be understood by analyzing the frequency spectrum in more detail. The resonance frequencies respectively wave numbers for TM-modes with small $n_r$ can be approximated as
\begin{equation} \label{eq:wgm_app} \Re{k_{m, n_r}} = \frac{m}{n R} + \frac{x_{n_r}}{nR} \left( \frac{m}{2} \right)^{1/3} - \frac{1}{R \sqrt{n^2 - 1}} \, , \end{equation}
where $x_j$ is the modulus of the $j$th zero of the Airy function $\mathrm{Ai}(x)$ \cite{Dubertrand2008}. Therefore, the resonance spacing between modes with the same radial quantum number $n_r$ is
\begin{equation} \Delta k = \frac{1}{n R} + \frac{x_{n_r}}{6 n R} \left( \frac{m}{2} \right)^{-2/3} \, , \end{equation}
and the smallest resonance spacing $(\Delta k)_{\mathrm{min}}$ is determined by the highest azimuthal quantum number $m$ or resonance frequency $\fmax$, respectively. Since the resonance spacing is almost constant, only one peak is expected in the FT of the spectrum, and its position is related to $(\Delta k)_{\mathrm{min}}$. These observations are also made for the other two measured families [see \reffig{subfig:Lspekt_Tefcirc_single_nr}].

\begin{figure*}[t]
\includegraphics[width = 16 cm]{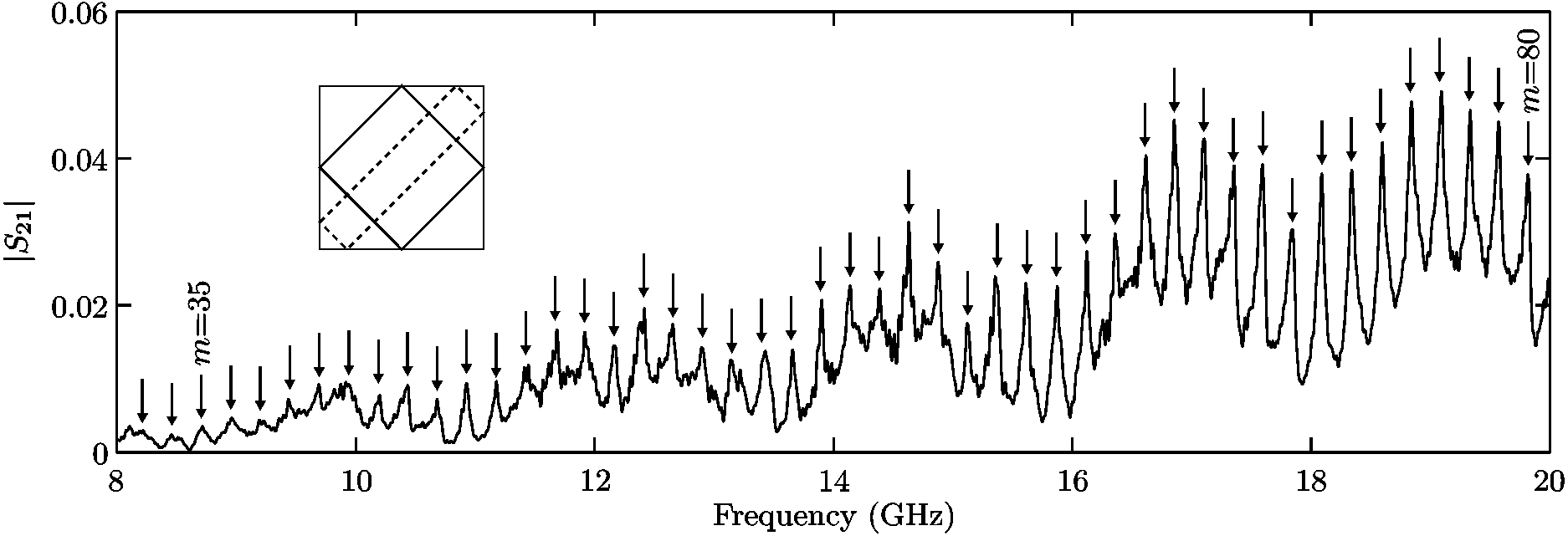}
\caption{\label{fig:spek_Tefsquare}Frequency spectrum of the Teflon square. The series of equidistant resonances atop the slowly oscillating background are superscarred states localized on a family of POs including the diamond orbit (shown as inset). The arrows indicate the computed resonance frequencies of the superscarred states, and the corresponding quantum numbers $m = 35$ and $m = 80$ are indicated for two examples, respectively.}
\end{figure*}

As a next step, we consider the length spectrum taking into account several families of resonances in \reffig{subfig:Lspekt_Tefcirc_several_nr}. The combination of just two families (solid length spectrum) is enough to obtain multiple peaks, whose positions are indeed very close to the lengths of the POs. This effect can be explained as an interference between the different families (see Appendix \ref{sec:fam_int}). Moreover, the resonances with $n_r = 1$ and $2$ contribute mainly to the high-order polygons such as the hexagon and the heptagon, whose lengths are in the region where the corresponding single-family length spectra [full and dashed lines in \reffig{subfig:Lspekt_Tefcirc_single_nr}] are maximal. The resonances with $n_r = 3$ [added to obtain the dashed length spectrum in \reffig{subfig:Lspekt_Tefcirc_several_nr}] mainly contribute to the pentagon orbit, which again lies in the length region where the corresponding single-family length spectrum is maximal, and the few remaining resonances (added to obtain the dotted length spectrum) provide contributions to the pentagon and square orbits. So there is indeed a connection between the trajectories to which the different resonances correspond and the POs in the length spectrum to which they contribute, and the deviations between the peak amplitudes of the measured and calculated length spectra are larger for the square and pentagon orbits (see \reffig{fig:Lspekt_Tefcirc_all}) because the most long-lived states correspond to the higher-order polygons. These results were also confirmed by investigations with a complete, calculated spectrum (not shown here). Additional losses due to, e.g., absorption in the Teflon material further reduce the amplitudes of the experimental length spectrum, but this is only a secondary effect due to the generally high quality factors of the Teflon circle.

\section{\label{sec:teflonsquare}Square Teflon resonator}

\begin{figure*}[t]
\includegraphics[width = 16 cm]{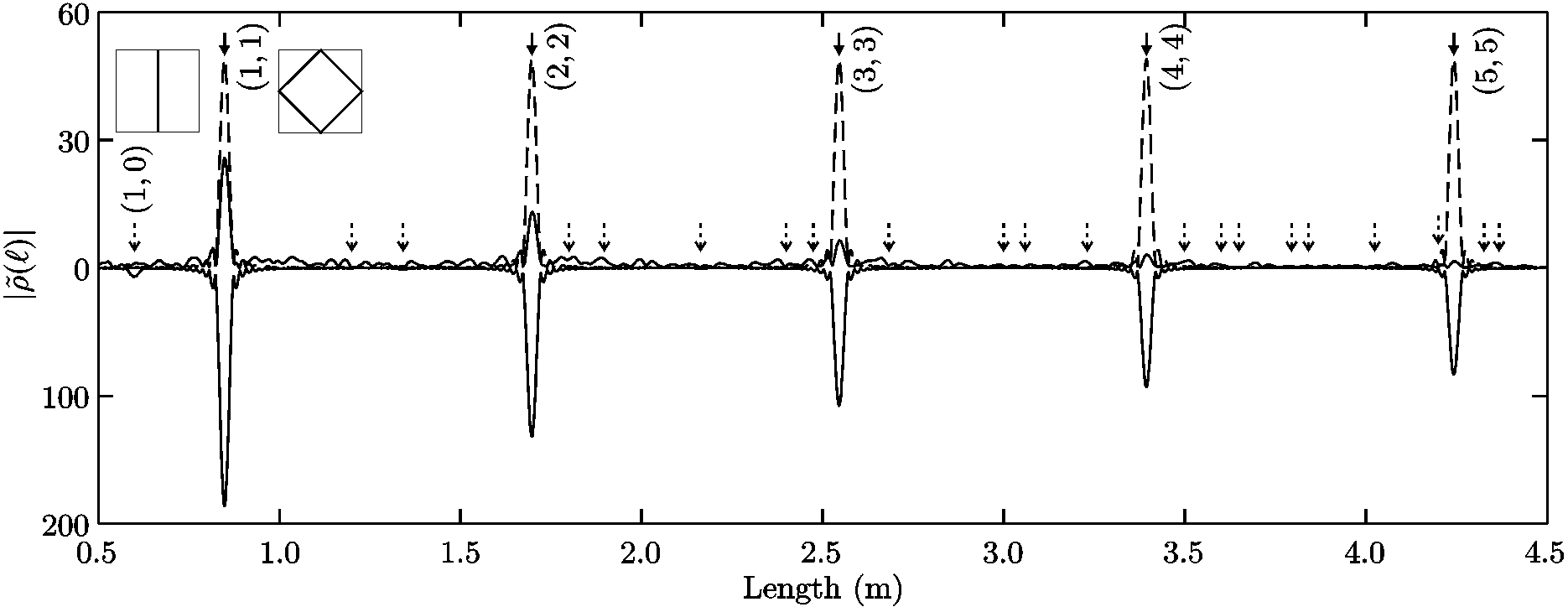}
\caption{\label{fig:lspekt_Tefsquare}Length spectrum of the Teflon square. The top graph shows the experimental length spectrum (solid line) and the length spectrum for a single family of superscars (dashed line), and the bottom graph the semiclassical expression. Note the different scales of top and bottom graphs. The arrows denote the lengths of the POs in the square (dashed arrows for orbits not contained by TIR), and the indices $(n_x, n_y)$ denote the type of orbit. Two examples of POs  are shown as insets [Fabry-Perot orbit (left) and diamond orbit (right)]. The experimental length spectrum features only the diamond orbit and its repetitions, which are the only POs confined by TIR.}
\end{figure*}

The second resonator investigated is a square disk made of Teflon (called the Teflon square in the following) with side length \mbox{$a = 300.0$ mm}, thickness \mbox{$d = 5.1$ mm} and index of refraction $n = 1.430 \pm 0.001$, which corresponds to $\alphacrit = 44.4 ^\circ$. Its frequency spectrum is shown in \reffig{fig:spek_Tefsquare}, with a frequency of \mbox{$10$ GHz} corresponding to \mbox{$ka = 62.9$}. The spectrum features a single family of broad, equidistant resonances atop an oscillating background. This background results from direct transmission between the two antennas. The quality factors of the resonances are in the range of $Q = 100 \-- 500$, which is an order of magnitude lower than for the Teflon circle due to the larger radiation losses of the Teflon square. The resonances can be explained as superscarred states \cite{Lebental2007} localized on the family of the diamond PO (inset of \reffig{fig:spek_Tefsquare}). As the angle of incidence of the diamond orbit, $\alpha_\po = 45 ^\circ$, is very close to the critical angle, the radiation losses are large. The resonance frequencies can be approximated \cite{Lebental2007} up to $O \left( \frac{1}{m} \right)$ as
\begin{equation} \label{eq:SSfreq} n L \, \Re{k_m} = \pi m + 2 i \ln[r(45^\circ)] = \pi m + 4 \delta \, , \end{equation}
with $L = \sqrt{2} \, a$ being half the length of the diamond PO and $m$ being the longitudinal quantum number of the superscar. The phase $\delta = \arctan{(\sqrt{1 - 2 / n^2})}$ is related to the Fresnel reflection coefficient via $r(45^\circ) = \exp{(-2 i \delta)}$, and the term $4 \delta$ accounts for the reflections at the dielectric boundaries of the square. The resonance frequencies computed from \refeq{eq:SSfreq} are indicated by the arrows in \reffig{fig:spek_Tefsquare} and show good agreement with the measured spectrum. Apparently, the observed resonances are superscarred states with longitudinal quantum numbers in the regime of $m \approx 35 \-- 80$ and first transverse excitation. States with higher transverse excitation are not observed. All in all, $49$ resonances were counted up to \mbox{$20.5$ GHz}, that is only $2 \%$ of the approximately $2220$ resonances expected according to Weyl's law.

\begin{figure*}[tb]
\subfigure[]{
	\includegraphics[height = 5.2 cm]{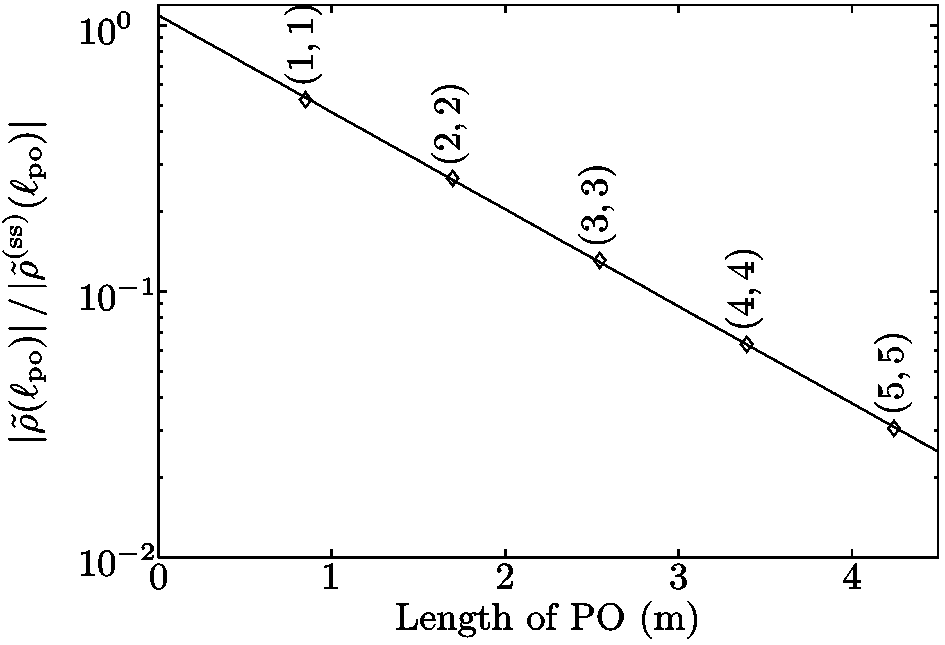}
	\label{subfig:RelAmp_Tefsquare}
}
\hspace{0.5 cm}
\subfigure[]{
	\includegraphics[height = 5.2 cm]{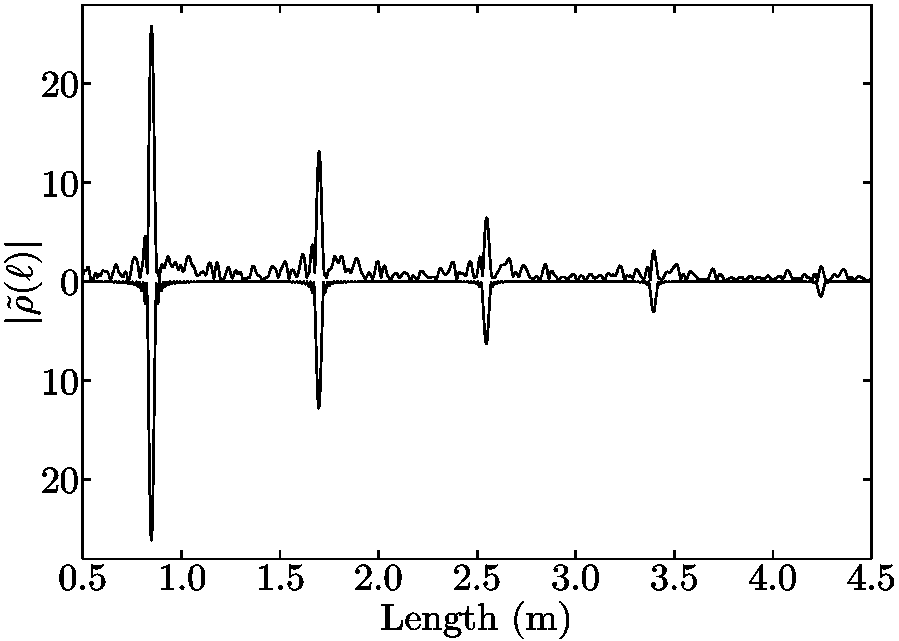}
	\label{subfig:lspekt_Tefsquare_fit}
}	
\caption{\label{fig:RelAmp_Tefsquare}\subref{subfig:RelAmp_Tefsquare} Ratio between the peak amplitudes of the experimental and the superscar length spectra. The diamonds are the relative amplitudes for the indicated POs, and the full line is an exponential fit of \refeq{eq:exp_fit} whose slope corresponds to a resonance width of \mbox{$\Gamma = 56.1$ MHz}. \subref{subfig:lspekt_Tefsquare_fit} Experimental length spectrum (top graph) and superscar length spectrum multiplied with the exponential fit (bottom graph). The two curves show excellent agreement.}
\end{figure*}

In \reffig{fig:lspekt_Tefsquare}, the experimental length spectrum (solid line in the top graph) and the semiclassical expression $|\rhotscl{(\ell)}|$ (bottom graph) for the Teflon square are shown. The arrows indicate the lengths of the different POs in the square, and dashed arrows are used for those POs not contained by TIR. The indices $(n_x, n_y)$ indicate half the number of bounces of the PO in $x$ and $y$ directions; see insets in \reffig{fig:lspekt_Tefsquare}, for examples. Details for the semiclassical amplitudes of the POs are given in Appendix \ref{sec:po_square}. Only the $(1, 1)$ orbit (diamond) and its repetitions are contained by TIR, and the experimental length spectrum only shows peaks at these lengths in agreement with the semiclassical prediction. Thus, the very simple structure of the frequency spectrum can be attributed to the single PO (plus its repetitions) contained by TIR. The trace formula also predicts a small peak at the length of the $(1, 0)$ orbit (Fabry-Perot orbit; left inset of \reffig{fig:lspekt_Tefsquare}), which however is not observed experimentally. The same applies for all other orbits not contained by TIR (indicated by the dashed arrows). The peak amplitudes of the experimental length spectrum are, however, less than $15 \%$ of the semiclassical prediction. Moreover, they decay exponentially with the number $\mu$ of repetitions, while the semiclassical amplitudes decay algebraically with $1 / \sqrt{\mu}$ (see Appendix \ref{sec:po_square}). This is related to the facts that, first, only a single family of superscarred resonances is observed in the spectrum, i.e., the resonator essentially behaves like a one-dimensional system, and, second, that the resonances have a finite lifetime. Starting with a single family of modes with resonance frequencies given by \refeq{eq:SSfreq}, one obtains
\begin{equation} \label{eq:rho_SS} \rhofSS(k) = \frac{2 n L}{\pi} \sum \limits_{\mu = 1}^{\infty} \cos{(2 \mu n L k - 8 \mu \delta)} \end{equation}
as the semiclassical density of states \cite{Brack2003}. The FT of $\rhofSS(k)$, i.e., the length spectrum for a single family of superscars ($\mathrm{ss}$), is plotted as dashed line in the top graph in \reffig{fig:lspekt_Tefsquare} and has a constant peak amplitude for all repetitions $\mu$ of the diamond orbit. The agreement with the experimental length spectrum is better. However, \refeq{eq:rho_SS} does not take into account that the system is open; in fact only the real part of $k$, i.e., the resonance frequencies, not the widths $\Gamma$ are considered. The ratio of the peak amplitudes of the experimental length spectrum (full line) and the superscar length spectrum (dashed line) is shown in \reffig{subfig:RelAmp_Tefsquare} with respect to the lengths $\lpo$ of the (repeated) diamond orbit. The solid line is a fit of the function 
\begin{equation} \label{eq:exp_fit} A_0 \exp{(-n \lpo \pi \Gamma / c)} \, , \end{equation}
with $A_0 = 1.09$ and \mbox{$\Gamma = 56.1$ MHz}, which roughly matches the widths of the measured resonances. Indeed, \reffig{subfig:lspekt_Tefsquare_fit} shows that the experimental length spectrum (top graph) and the FT of $\rhofSS(k)$ multiplied with the fitted exponential decay corresponding to the finite lifetimes of the resonances (lower graph) agree with high precision. \newline
The example of the Teflon square demonstrates that a dielectric billiard with just a single dominant PO (the diamond orbit in this case) may act as an effectively one-dimensional system. Note that for a larger index of refraction or for higher frequencies, further families of superscarred states (i.e., with higher transverse excitation) might become visible in the frequency spectrum. Furthermore, the Teflon square is an example where a single family of resonances is directly related to one PO, which is in contrast to the case of the dielectric circle billiard (see \reffig{fig:Lspekt_Tefcirc_nr1} and corresponding text). The reason for this is that the resonances in the dielectric square are (super)scarred states, while those in the circle billiard are not. It appears that the relation of a single resonance family to POs depends on the specific case and needs careful analysis.

\section{\label{sec:aluminasquare}Square alumina resonator}

\begin{figure*}[tb]
\includegraphics[width = 16 cm]{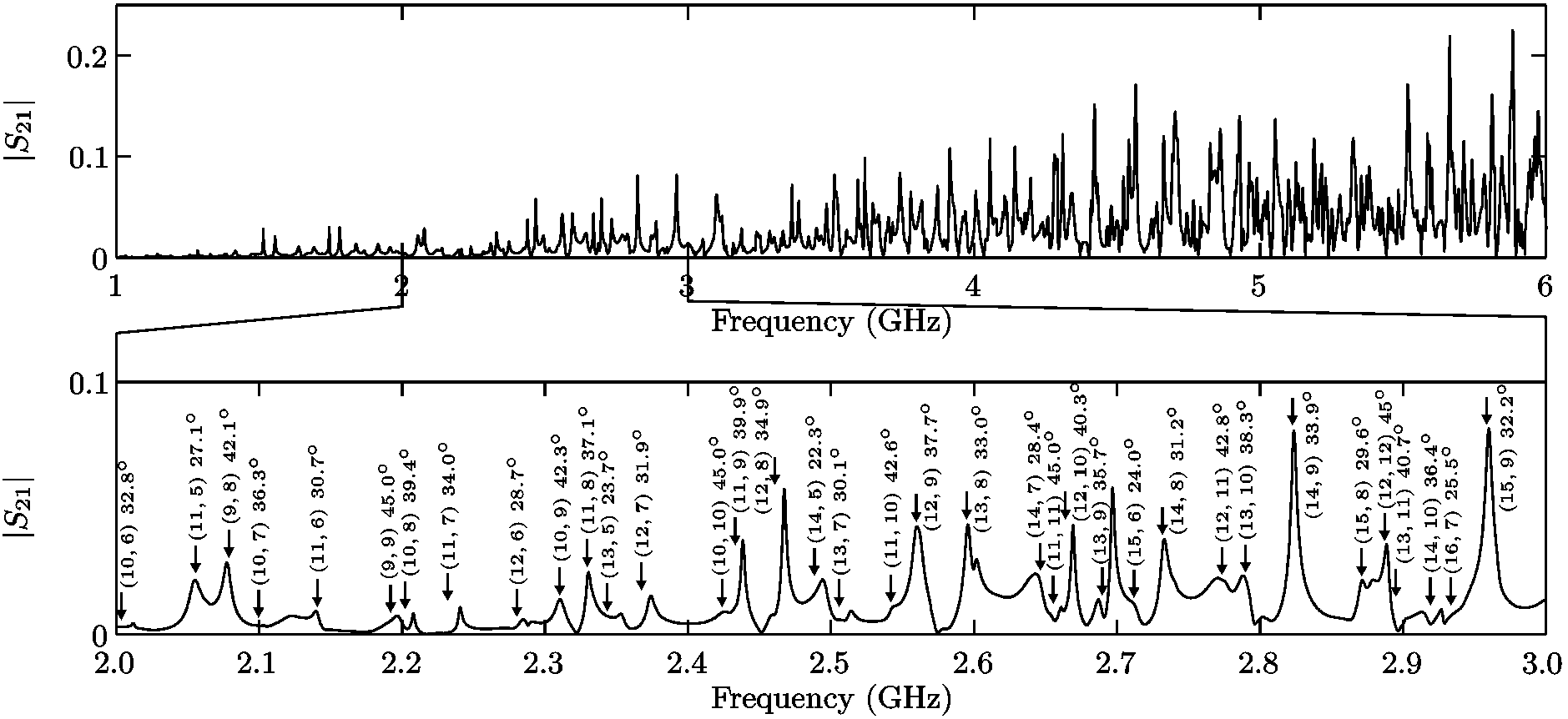}
\caption{\label{fig:spect_alsquare}Frequency spectrum of the alumina square. In contrast to the cases of the Teflon circle and square, the structure of the spectrum seems irregular, and it cannot be divided into one or more families of equidistant resonances. The lower graph shows the part of the spectrum from $2$ to $3$ GHz. The arrows indicate the resonance frequencies computed according to a generalized superscar model, and the quantum numbers $(m_x, m_y)$ and the corresponding angles of incidence $\tilde{\chi}$ are given for each. The computed resonance frequencies agree well with the measured spectrum in most cases.}
\end{figure*}

\begin{figure*}[tb]
\includegraphics[width = 16 cm]{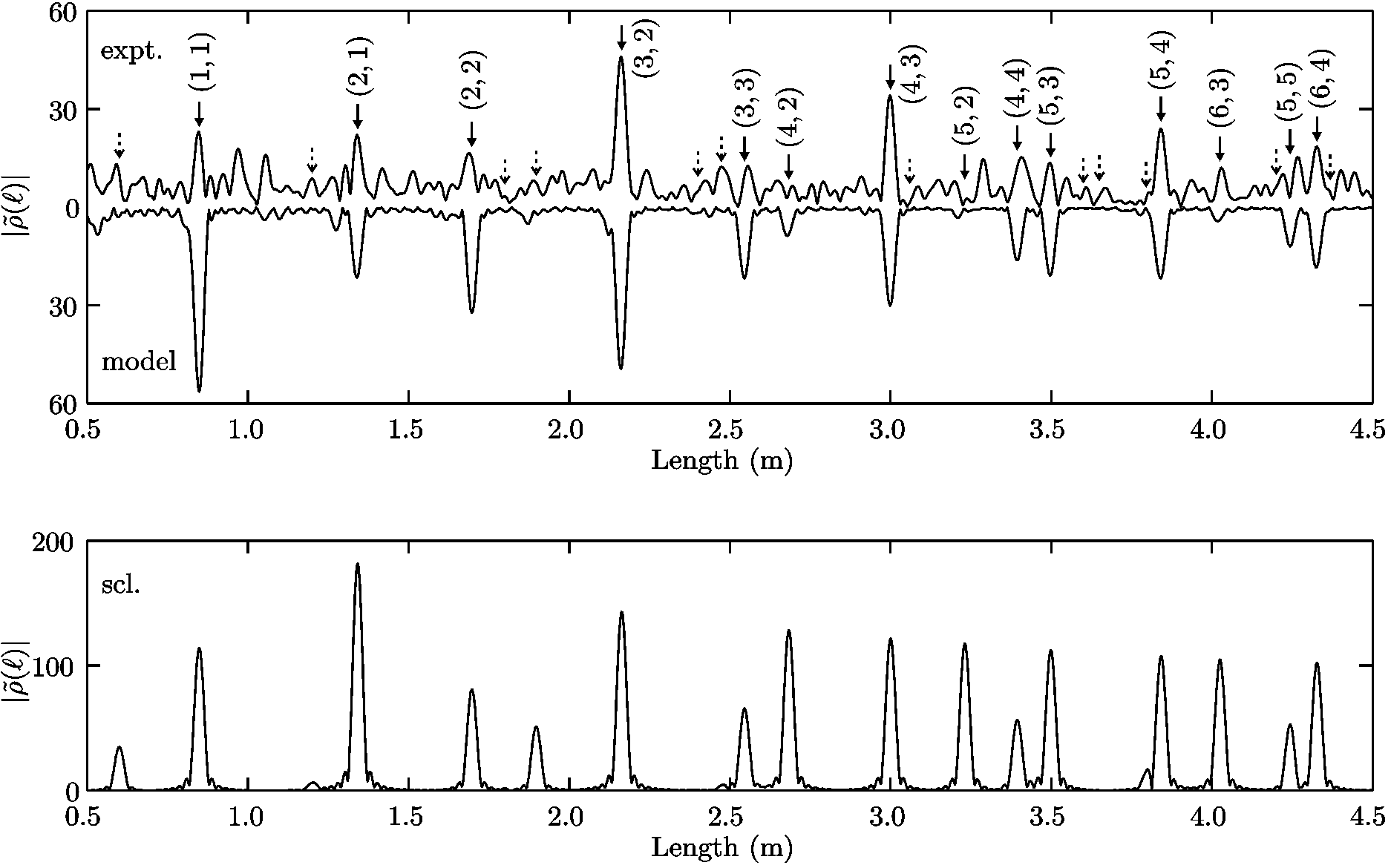}
\caption{\label{fig:lspekt_alsquare}Length spectrum of the alumina square. The upright graph in the top panel is the experimental length spectrum, the inverse graph is based on the generalized superscar model, and the bottom panel shows the semiclassical expression. Note the different scales of top and bottom panels. The lengths of the POs are indicated by arrows, with solid arrows for those POs confined by TIR and dashed arrows for those not confined. The indices $(n_x, n_y)$ characterizing the confined POs are also given. There are several peaks corresponding to POs in the experimental length spectrum, but most of them hardly stand out above the noise level, and their peak amplitudes are much lower than semiclassically predicted. The inverse graph in the top panel is the length spectrum for a set of modes computed according to the generalized superscar model. Only modes with $\tilde{\chi} \geq 28^\circ$ were taken into account in this set, and no possible degeneracies. This length spectrum reproduces the experimental one well except for the cases of the $(1, 1)$ and $(2, 2)$ orbit.}
\end{figure*}

\begin{figure}
\includegraphics[width = 8.6 cm]{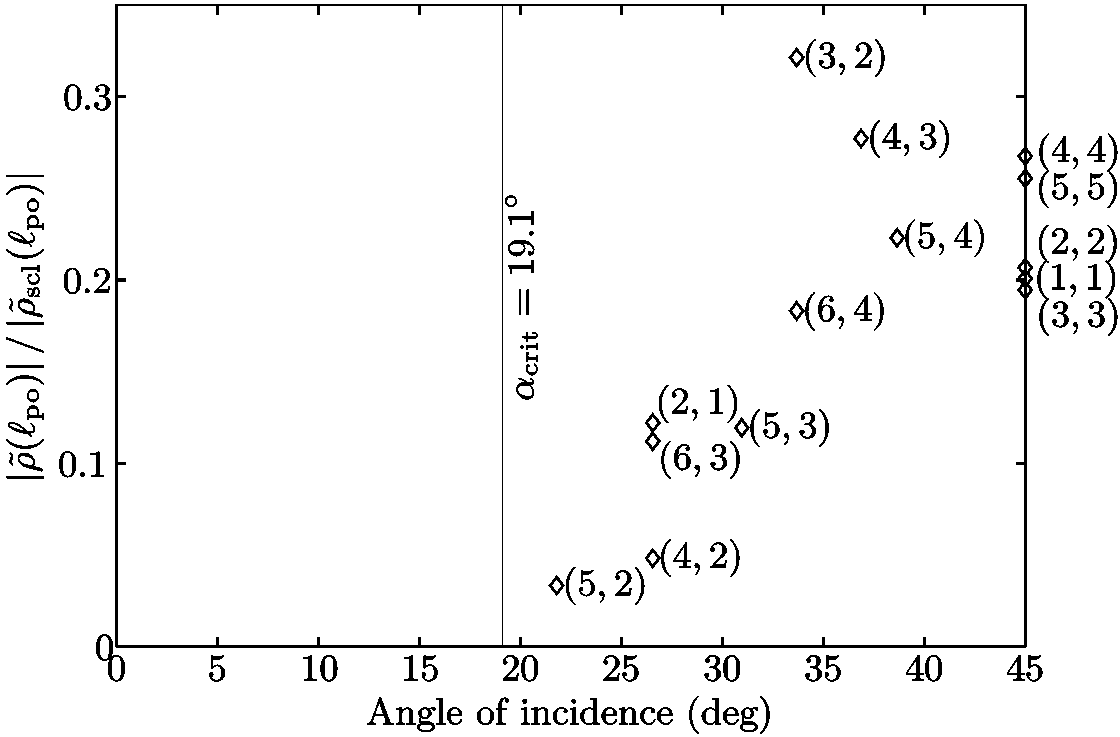}
\caption{\label{fig:RelAmp_alsquare}The ratio between the peak amplitudes of the experimental and semiclassical length spectra with respect to the angle of incidence of the corresponding PO. The indices $(n_x, n_y)$ of the POs are indicated. The vertical line signifies the critical angle for TIR. The relative amplitudes for POs close to the critical angle are smaller compared to those of other POs.}
\end{figure}

The third resonator investigated is a square disk made of alumina (Morgan Advanced Ceramics) with side length \mbox{$a = 300.0$ mm}, index of refraction $n = 3.050 \pm 0.008$, and thickness \mbox{$d = 8.3$ mm}. The critical angle is $\alphacrit = 19.1^\circ$. A measured frequency spectrum of the resonator (called the alumina square in the following) is shown in \reffig{fig:spect_alsquare}, with \mbox{$5$ GHz} corresponding to \mbox{$ka = 31.4$}. The spectrum features many sharp resonances, and $212$ resonances were identified in the range of $1.4 - 6.1$ GHz (without taking into account possible degeneracies), compared to $1035$ resonances expected according to Weyl's law. The quality factors are in the regime of $Q = 200 \-- 2000$. The radiation losses are generally smaller than for the Teflon square due to the higher index of refraction, but the absorption losses in the alumina material are somewhat larger than in the Teflon material. In contrast to the case of the Teflon square, the spectrum shows no obvious structure of equidistant resonances, so the simple superscar model from \refsec{sec:teflonsquare} cannot be applied here because it takes into account only one family of periodic orbits with an angle of incidence of $45^\circ$. However, it can be generalized in a simple way \cite{Bogomolny2009a}. We assume a ray with wave vector $(k_x, k_y)$ traveling in the square, where the $x$ and $y$ axes are parallel to the sides of the square, and the resonance condition after one round trip is 
\begin{equation} \begin{array}{c} e^{2 i a k_x} r^2(\chi) = 1 \, ,  \\ e^{2 i a k_y} r^2(\pi/2 - \chi) = 1 \, , \end{array} \end{equation}
with $r(\chi)$ being the Fresnel reflection coefficient and $\chi$ being the angle of incidence on the vertical sides. An approximate solution is
\begin{equation} \label{eq:SSfreq_gen} \begin{array}{rcl} k_x a & = & m_x \pi + i \ln{[r(\chi)]} \, , \\ k_y a & = & m_y \pi + i \ln{[r(\pi/2 - \chi)]} \, , \end{array} \end{equation}
where $(m_x, m_y)$ are the $x$ and $y$ quantum numbers and the angle of incidence is approximated as $\chi = \arctan{(m_y / m_x)}$. The wave number is $k = \sqrt{k_x^2 + k_y^2} /n$. This simple semiclassical approximation can be regarded as a generalized superscar model. A similar model was proposed in \cite{Che2010a}. The case $m_x \approx m_y$, $\chi \approx 45^\circ$ corresponds to the simple superscar model used in \refsec{sec:teflonsquare} and resonance frequencies as in \refeq{eq:SSfreq} are obtained. The resonance frequencies computed according to \refeq{eq:SSfreq_gen} are indicated by arrows in the magnified part of the spectrum (lower graph in \reffig{fig:spect_alsquare}). The quantum numbers are given together with $\tilde{\chi} = \arctan{(k_y / k_x)} \approx \chi$, which is the angle of incidence on the vertical sides corresponding to the wave vector $(k_x, k_y)$. In most cases, the measured and the computed resonance frequencies agree very well, and only in some cases they slightly deviate. Moreover, there are very few cases (e.g., at $2.7$ GHz) where we cannot find a clear correspondence between a measured resonance and a computed mode. Only modes which are found in the measured spectrum are indicated. Modes with quantum numbers $m_x \neq m_y$ are doubly degenerate. It should be noted that \refeq{eq:SSfreq_gen} predicts complex resonance frequencies [$\Im{k} < 0$] for $\tilde{\chi} < \alphacrit$, that is, losses due to refractive escape, and lossless modes [$\Im{k} = 0$] for $\tilde{\chi} > \alphacrit$. Note that in the latter case other radiative loss mechanisms, e.g., due to the corners \cite{Wiersig2003}, are not accounted for. Accordingly, the model does not provide a complete description for the widths of these modes. All modes found in the measured spectrum have $\tilde{\chi} > \alphacrit$. One would intuitively expect that the losses increase with $\tilde{\chi}$ approaching $\alphacrit = 19.1^\circ$ starting with $\tilde{\chi} = 45^\circ$. Accordingly, the spectrum should be dominated by an equidistant series of resonances corresponding to $\tilde{\chi} \approx 45^\circ$, as was observed in the case of the Teflon square. However, a closer inspection of the spectrum shows no clear correlation between $\tilde{\chi}$ and the widths or amplitudes of the resonances. We attribute this effect to interaction between superscar and background states \cite{Aberg2008}. This would also explain why the computed resonance frequencies agree precisely with the measured ones only in some cases. Preliminary numerical computations confirm this interpretation of the effect, but it is not yet completely understood and will be the subject of further investigations. In summary, the generalized superscar model explains the general structure of the spectrum well, but not all of its details like, e.g., the resonance widths. \newline
The length spectrum for the alumina square is shown in \reffig{fig:lspekt_alsquare}. The experimental length spectrum (upright graph in the top panel) features several peaks corresponding to POs, but most of the peaks hardly stand out above the noise level. The peak amplitudes of the experimental length spectrum are below $35 \%$ of those expected semiclassically (bottom panel), which is far less than for the Teflon circle, even though $20 \%$ of the total number of resonances was found in the case of the alumina square as compared to $10 \%$ in the case of the Teflon circle. A similar proportion between the number of resonances and the peak amplitudes was also found in numerical calculations for the alumina square. The reason why a larger percentage of observed resonances in the case of the alumina square results in smaller peak amplitudes compared to the case of the Teflon circle is not understood, but we surmise that this is related to qualitative differences in the distribution of resonance widths for the circle and the square billiard. No peaks corresponding to POs not confined by TIR (indicated by dashed arrows) can be identified, as for the Teflon circle and square, respectively. For the observed ones the ratio of the experimental and the semiclassical amplitudes varies significantly. This is illustrated in \reffig{fig:RelAmp_alsquare}, where it is plotted with respect to the angle of incidence $\alpha_\po$. Every trajectory in the square is characterized by two different angles of incidence (see Appendix \ref{sec:po_square}), of which the smaller one is used here. As in the case of the Teflon circle, even though they are still confined by TIR, POs with an angle of incidence close to the critical angle have significantly smaller relative peak amplitudes compared to those with $\alpha_\po$ much larger than $\alphacrit$. Finally, we compare the experimental length spectrum to the generalized superscar model introduced above. To this end, we calculate a spectrum according to \refeq{eq:SSfreq_gen}, but omit all modes with angle $\tilde{\chi}$ below a certain cutoff angle $\chico$, and include only one of the degenerate doublet modes since possible degeneracies are not taken into account in the experimental length spectrum either. The cutoff angle is chosen as $\chico = 28^\circ$ to ensure that the number of resonances ($215$) obtained from $1.4 - 6.1$ GHz is similar to that in the experiment ($212$). The lower part of \reffig{fig:spect_alsquare} shows that indeed almost all measured resonances correspond to a $\tilde{\chi} > \chico$. Note also that the relative amplitudes in \reffig{fig:RelAmp_alsquare} start to decrease at a similar angle. Since the generalized superscar model does not provide a description of the resonance widths in the cases considered here, we assume a uniform resonance width of $\Gamma = 8$ MHz as obtained from an estimate of the typical resonance width found in the measured spectrum. The resulting length spectrum is shown as an inverse graph in the top panel of \reffig{fig:lspekt_alsquare}. It agrees well with the experimental length spectrum except for the cases of the $(1, 1)$ and $(2, 2)$ orbits, for which the amplitudes are significantly larger than those of the experimental length spectrum. These deviations cannot be explained yet. Nevertheless, even though the generalized superscar model cannot reproduce all details of the frequency and length spectra of the alumina square, most of the experimental findings are well described by this simple semiclassical approximation if taking into account only the resonances which are observed in the experiment. It should be noted that the semiclassical trace formula [\refeq{eq:rho_scl}] for the dielectric square can be deduced from \refeq{eq:SSfreq_gen} when all of the modes are considered \cite{Bogomolny2009a}. \newline
The case of the alumina square demonstrates again that the long-lived, observable resonances contribute mainly to those POs with angle of incidence far away from the critical angle, indicating that corrections to the semiclassical trace formula must be taken into account for POs close to it. Furthermore, the larger index of refraction and thus greater number of POs contained by TIR are reflected in a more complicated structure of the frequency spectrum. In conclusion, not only the shape of a dielectric billiard, but also its index of refraction, i.e., the critical angle for TIR, drastically influences the structure of its frequency and length spectra.

\section{\label{sec:conc}Conclusions}
We have measured the frequency spectra of three 2D dielectric microwave resonators with different shapes and indices of refraction. The length spectra were determined from the measured data and compared to the predictions of the trace formula proposed in \cite{Bogomolny2008}. Even though the number of resonances identified in the measured spectra is only a fraction between $2 \%$ and $20 \%$ of the total number of resonances, the experimental length spectra show peaks connected with the classical POs as predicted, but the amplitudes of the peaks are smaller than the semiclassically expected ones. The differences between experimental and semiclassical peak amplitudes depend strongly on the angles of incidence of the PO: POs which are not contained by TIR are not observed in the experimental length spectra, even if a finite amplitude is predicted semiclassically. This was the case for all of the passive resonators investigated here, but there are indications that they may appear in the length spectra of strongly pumped microlasers \cite{Lebental2008}. Furthermore, even POs contained by TIR have much smaller amplitudes if their angles of incidence are close to the critical angle. Apparently, there are two different effects. First, the strict semiclassical approximation derived in the limit $k \rightarrow \infty$, when applied for large but finite $k$, requires modifications which are especially important close to the critical angle. From the soluble case of the disk it follows that the reflection coefficient for such scattering is noticeably smaller than that predicted by the usual Fresnel formulas \cite{Hentschel2002a, Hentschel2009}. Second, it appears that long-lived resonances that can be observed experimentally mainly correspond to orbits which are not only confined by TIR, but also have an angle of incidence much larger than the critical angle. The examples of the Teflon and alumina square billiards further demonstrate that there is a connection between the number of POs contributing to the length spectrum and the complexity of the frequency spectrum. Therefore, it could prove useful for the design of resonators to consider the type and number of POs confined for a certain billiard geometry and index of refraction. In general, the systematics of observed, long-lived and not observed, short-lived resonances must be taken into account for a full understanding of the experimental length spectra. Special care must be taken in the case of spectra with only one or two families of resonances. In general, there is no direct relation to certain POs. Only the case of (super)scarred states may be an exception to this rule, as the case of the Teflon square billiard demonstrates. \newline
In conclusion, we have demonstrated that the length spectra of dielectric resonators may be described by the trace formula, offering another tool to understand the correspondence between the ray- and wave-pictures of these devices. An advantage of the trace formula is that it only needs the spectrum, but not the field distributions as an input. On the other hand, the trace formula just gives information about a group of resonances in general, but not on the individual resonances, and at least a few dozens of resonances are needed in order to apply it. So far, the trace formula has merely been tested for passive systems with regular classical dynamics. Systems with chaotic or mixed dynamics or with an active medium (like microlasers) pose interesting future problems.

\begin{acknowledgments}
The authors are grateful to R. Dubertrand for providing numerical calculations for the square alumina resonator. F. S. acknowledges support from the Deutsche Telekom Foundation, and P. O. I. from the DAAD. This work was supported by the DFG within the Sonderforschungsbereich 634.
\end{acknowledgments}

\appendix

\section{\label{sec:po_circ}The periodic orbits in the dielectric circle billiard}
The POs in the circle billiard are characterized by their period and rotation numbers $(q, \eta)$, where $q$ is the number of reflections at the boundary and $\eta$ is the number of turns around the center. Their lengths are
\begin{equation} \lpo(q, \eta) = 2 R q \, \sin{(\eta \pi / q)} \, , \end{equation}
and the amplitudes and phases appearing in \refeq{eq:rho_scl} are 
\begin{equation} B_\po = \frac{A_\po}{\sqrt{\lpo}} \sqrt{2} \, f_\po \, , \, \mathrm{with} \, f_\po = \left\{ \begin{array}{ccl} 1 & , & q = 2 \eta \\ 2 & , & \mathrm{otherwise,} \end{array} \right. \end{equation}
\begin{equation} \varphi_\po = \frac{\pi}{4} - q \frac{\pi}{2} + q \arg{[r(\alpha_\po)]} \, . \end{equation}
Here, $r(\alpha)$ is the Fresnel reflection coefficient for angle of incidence $\alpha$ with respect to the surface normal and electric field perpendicular to the plane of incidence (TM polarization), and $\alpha_\po = \pi / 2 - \eta \pi / q$ is the angle of incidence of the PO. The area of the billiard covered by (the family of) the $(q, \eta)$ orbit(s) is $A_\po = \pi R^2 \sin^2{(\eta \pi / q)}$. These terms are identical to the case of a hard-walled circle billiard if $r(\alpha_\po) = -1$ (see, e.g., \cite{Sieber1995, Brack2003}). The amplitude $B_\po \propto \sin^{3/2}{(\eta \pi / q)}$, so $B_\po \rightarrow 0$ for $q/\eta \rightarrow \infty$. Therefore, the contribution to the length spectrum of the infinitely many orbits with length close to the circumference is vanishingly small.

\section{\label{sec:fam_int}Length spectrum for two resonance families of the dielectric circle billiard}
Beginning with expression (\ref{eq:wgm_app}) for the resonance frequencies of the WGMs, a trace formula for the spectrum with just one or two resonance families can be derived. The (approximate) inverse of \refeq{eq:wgm_app} is
\begin{equation} \label{eq:wgm_app_inv} m = A k - B x_{n_r} k^{1/3} + \frac{n}{\sqrt{n^2 - 1}} \, , \end{equation}
with $A = n R$, $B = (n R / 2)^{1/3}$, and $x_j$ being the modulus of the $j$th zero of the Airy function. Following Chap.\ 3.2 of \cite{Brack2003} we obtain
\begin{eqnarray} \rhofnr{1}(k) = && \, 4 \left (A - \frac{1}{3} B x_{n_r} k^{-2/3} \right) \nonumber \\* && \times \sum \limits_{\eta = 1}^{\infty} \cos{\left[ 2 \pi \eta (A k - B x_{n_r} k^{1/3}) \right]} \end{eqnarray}
as the fluctuating part of the density of states for a single resonance family. A factor of $2$ was introduced to account for the degeneracy of the resonances and the last term in \refeq{eq:wgm_app_inv} was neglected. Adding up two of these terms for different radial quantum numbers and applying a trigonometric addition theorem results in
\begin{widetext}
\begin{equation} \rhofnr{2}(k) \approx 8 \left( A - \frac{1}{3} B \bar{x} k^{-2/3} \right) \sum \limits_{\eta = 1}^{\infty} \cos{\left[ 2 \pi \eta (A k - B \bar{x} k^{1/3}) \right]} \cos{\left[ \pi \eta B \Delta x k^{1/3} \right]} \, , \end{equation}
\end{widetext}
for two resonance families (e.g., $n_r = 1$ and $2$) with $\bar{x} = (x_2 + x_1)/2$ and $\Delta x = x_2 - x_1$. The first cosine term is a fast oscillating function, while the other terms are only slowly varying in $k$, and the FT of $\rhofnr{2}$ is computed using the stationary phase approximation. The result is
\begin{eqnarray} \label{eq:rhot_2nr} \rhotnr{2}(\ell) = && \frac{2 \ell n}{\eta} \sqrt{\frac{3}{\pi}} \sum \limits_{\eta = 1}^{\infty} e^{i \varphi_\eta} \sqrt[4]{\frac{(\frac{2}{3} \pi \eta B \bar{x})^3}{(2 \pi \eta A - \ell n)^5}} \nonumber \\* && \times \cos{\left[ \pi \eta B \Delta x \sqrt{\frac{\frac{2}{3} \pi \eta B \bar{x}}{2 \pi \eta A - \ell n}} \right]} \, . \end{eqnarray}
The details on the stationary phase $\varphi_\eta$ are omitted. It should be noted that this expression (for each $\eta$ individually) is only valid if the stationary point lies within the integration interval, which is the case for $\ell \leq 2 \pi \eta R [1 - \frac{1}{3 n R} (n R / 2)^{1/3} \bar{x} \kmax^{-2/3}]$. For the same reason the positions of the maxima in \reffig{fig:Lspekt_Tefcirc_nr1} depend on $\fmax$. The modulus of the cosine term in \refeq{eq:rhot_2nr} is maximal if its argument equals $q \pi$, i.e., for
\begin{equation} \label{eq:lmax_2nr} \lmax(q, \eta) = 2 \pi R \eta \left[ 1 - \frac{\bar{x}}{6} \left( \frac{\eta \Delta x}{q} \right)^2 \right] \, . \end{equation}
The numerical values of $\lmax(q, \eta)$ are close to the lengths $\lpo(q, \eta)$ of the POs with $(q, \eta)$ for large $q / \eta$, as shown in \reftab{tab:lpo_vs_lmax}, which explains why the length spectrum for just two families of resonances in \reffig{subfig:Lspekt_Tefcirc_several_nr} features peaks at the lengths of these POs. Indeed, it can be shown that
\begin{equation} \lpo(q / \eta \rightarrow \infty) = 2 \pi R \eta \left[ 1 - \frac{1}{6} \left( \frac{\eta \pi}{q} \right)^2 \right] \, , \end{equation}
and that the zeros $x_j$ of the Airy function fulfill $\bar{x} (\Delta x)^2 \approx \pi^2$ for $\bar{x} = (x_{j+1} + x_j)/2$ and $\Delta x = x_{j+1} - x_j$. \newline \newline

\begin{table}[bt]
\caption{\label{tab:lpo_vs_lmax}Comparison of the lengths $\lpo$ of the periodic orbits in the Teflon circle with the length $\lmax$ according to \refeq{eq:lmax_2nr}. The values are computed for \mbox{$R = 274.9$ mm} and $\eta = 1$. The agreement is very good for large $q$, and reasonable even for smaller $q$.}
\begin{ruledtabular}
\begin{tabular}{rrrr}
$q$ & $\lpo(q, \eta = 1)$ (m) & $\lmax(q, \eta = 1)$ (m) & $\Delta \ell$ (m) \\
\hline
 3 & 1.4284 & 1.4126 & 0.0158 \\
 4 & 1.5551 & 1.5502 & 0.0049 \\
 5 & 1.6158 & 1.6140 & 0.0018 \\
 6 & 1.6494 & 1.6486 & 0.0008 \\
 7 & 1.6698 & 1.6694 & 0.0004 \\
 8 & 1.6832 & 1.6830 & 0.0002 \\
 9 & 1.6924 & 1.6923 & 0.0001 \\
10 & 1.6990 & 1.6989 & 0.0001 \\
\end{tabular}
\end{ruledtabular}
\end{table}

\section{\label{sec:po_square}The periodic orbits in the dielectric square billiard}
A PO traversing the square billiard $2 n_x$ ($2 n_y$) times in the $x$ ($y$) direction is denoted by the two indices $(n_x, n_y)$. The length of the PO is
\begin{equation} \lpo(n_x, n_y) = 2 a \sqrt{n_x^2 + n_y^2} \, , \end{equation}
where $a$ denotes the side length of the square. Since all families of POs cover the whole billiard area, the amplitudes and phases entering \refeq{eq:rho_scl} are
\begin{equation} B_\po = \frac{F_\po}{\sqrt{2}} \frac{a^2}{\sqrt{\lpo}} = \frac{F_\po}{2} \frac{a^{3/2}}{\sqrt[4]{n_x^2 + n_y^2}} \, , \end{equation}
with $F_\po = 2$ in the case of diamond and Fabry-Perot orbits and $F_\po = 4$ for all other POs \cite{Sieber1995}, and
\begin{equation} \varphi_\po = - \frac{\pi}{4} + \arg{(R_\po)} \, . \end{equation}
Like for all regular billiards, $B_\po \propto 1 / \sqrt{\lpo}$ and especially $B_\po \propto 1 / \sqrt{\mu}$, where $\mu$ is the number of repetitions of a PO. The total Fresnel coefficient is $R_\po = r^{2 n_x}(\alpha_\po) \, r^{2 n_y}(\alpha_\po')$, and the angles of incidence on the vertical and horizontal sidewalls are $\alpha_\po = \arctan{(n_y / n_x)}$ and $\alpha_\po' = \pi / 2 - \alpha_\po$, respectively.

\end{document}